\documentclass[aps,pra,twocolumn,showpacs,superscriptaddress,floatfix]{revtex4}

\usepackage{amsmath}
\usepackage{txfonts}
\usepackage{microtype}
\usepackage{graphicx}
\usepackage{color}
\usepackage{ulem}
\def\sout{\bgroup\markoverwith
  {\textcolor{red}{\rule[0.5ex]{2pt}{0.5pt}}}\ULon}

\graphicspath{{figures/}}

\begin{document}

\title{Spin effects in relativistic ionization with highly charged ions in super-strong laser fields}

\author{Michael \surname{Klaiber}}\thanks{\mbox{Corresponding author: klaiber@mpi-hd.mpg.de }}
\affiliation{Max-Planck-Institut f\"ur Kernphysik, Saupfercheckweg 1, 69117 Heidelberg, Germany}
\author{Enderalp \surname{ Yakaboylu}} 
\affiliation{Max-Planck-Institut f\"ur Kernphysik, Saupfercheckweg 1, 69117 Heidelberg, Germany}
\author{Carsten \surname{M\"uller }} 
\affiliation{Institut f\"ur Theoretische Physik I, Heinrich-Heine Universit\"at D\"usseldorf,
Universit\"atsstrasse 1, 40225 D\"usseldorf, Germany }
\author{Heiko \surname{Bauke}} 
\affiliation{Max-Planck-Institut f\"ur Kernphysik, Saupfercheckweg 1, 69117 Heidelberg, Germany}
\author{Gerhard G. \surname{Paulus}} 
\affiliation{Institut f\"ur Optik und Quantenelektronik, Friedrich-Schiller-Universit\"at, Max-Wien-Platz 1, 07743 Jena, Germany}
\affiliation{Helmholtz Institut, Fr\"obelstieg 13, 07743 Jena, Germany}
\author{Karen Z. \surname{Hatsagortsyan}} 
\affiliation{Max-Planck-Institut f\"ur Kernphysik, Saupfercheckweg 1, 69117 Heidelberg, Germany}

\date{\today}

\begin{abstract}

Spin effects in above-threshold ionization of hydrogenlike highly charged ions in super-strong laser fields are investigated. Spin-resolved ionization rates in the tunneling regime are calculated by employing two versions of a relativistic Coulomb-corrected strong-field approximation (SFA). An intuitive simple-man model is developed which explains the derived scaling laws for spin-flip and spin-asymmetry effects. The intuitive model as well as our ab initio numerical simulations support the analytical results for the spin effects obtained in the dressed SFA where the impact of the laser field on the electron spin evolution in the bound state is taken into account. In contrast, the standard SFA is shown to fail in reproducing spin effects at ionization even at a qualitative level. The anticipated spin-effects are expected to be measurable with modern laser techniques combined with an ion storage facility.

\end{abstract}


\maketitle
 
State-of-the-art laser technology provides field intensities in which a free electron can achieve relativistic energies \cite{RMP_2012} stimulating  the experimental investigation of the relativistic regime of laser-atom interaction in ultra-strong fields in a plasma environment \cite{Moore_1999,Chowdhury_2001,Dammasch_2001,Yamakawa_2003,Gubbini_2005,DiChiara_2008,Palaniyappan_2008,DiChiara_2010}. However, to investigate subtle features of the relativistic tunneling dynamics at strong field ionization, one may need to realize a pure collision of a single highly charged ion  with a strong pulse of laser radiation. While separately ultra-strong lasers and highly charged ions are available in many experimental labs, only in a few places both tools are brought together, e.g., at GSI in Darmstadt, Germany where a storage ring for highly charged ions is complemented with the petawatt laser-system  PHELIX \cite{PHELIX}, providing 
infrared radiation of 
up to $10^{20}$ W/cm$^2$ intensity. This offers a unique opportunity for the experimental investigation of the relativistic regime of above-threshold ionization (ATI).

The main workhorses for the analytical treatment in strong field physics, the strong field approximation (SFA) \cite{Reiss_1990,Reiss_1990b} and the imaginary-time method (ITM) \cite{Popov_1997,Mur_1998,Popov_2004,Milosevic_2002r1,Milosevic_2002r2}, have been applied for calculation of ionization rates in the relativistic regime.
The relativistic features of the momentum distribution of the ionized electron in ATI are well-known, see, e.g., \cite{Becker_2002,Klaiber_2005}. They are essentially determined by the electron continuum dynamics in the laser field. In contrast to that, there is no clear understanding of the electron spin dynamics in relativistic ATI. Some spin effects have been evaluated by ITM, adding a heuristic spin-action to the quasi-classical action function \cite{Popov_2004}. 
The spin-asymmetry at ionization with a circularly polarized laser field has been investigated in \cite{Faisal_2004} using the standard SFA (S-SFA). However, the main deficiency of the S-SFA is that the Coulomb field effect of the atomic core on the electron continuum dynamics as well as the laser field influence on the bound state dynamics are neglected. 
In particular, the latter may be of crucial relevance for the electron's spin dynamics.

Generally, spin effects in laser fields have been in the focus of theoretical attention for a long time, in particular, for free electron motion and scattering \cite{Bunkin_1972,Szymanowski_1997,Walser_1999,Walser_2000b,Panek_2002,Panek_2004,Ahrens_2012,DiPiazza_2010,Muller_2011,
Muller_2012}. In the relativistic laser-atom interaction spin effects have been shown to appear in the laser-driven bound electron dynamics \cite{Hu_1999,Hu_2001,Walser_2001,Walser_2002}.  The spin dynamics in nonsequential double ionization of helium has been considered in \cite{Bhattacharyya_2007,Bhattacharyya_2011}.

In this paper spin effects are investigated in the relativistic tunneling ionization process of highly charged hydrogenlike ions in super-strong laser fields of linear and circular polarization. 
Spin-resolved ionization probabilities are analytically calculated employing a relativistic Coulomb corrected dressed SFA (D-SFA). In the D-SFA, in contrast to S-SFA,  
the laser field driven electron spin dynamics in the bound state is accounted for, using a nonstandard partition of the Hamiltonian within the SFA formalism \cite{Faisal_2007a,Faisal_2007b}. 
Asymptotic scaling laws for spin-flip and spin-asymmetry parameters are deduced. An intuitive simple-man's picture for spin effects is proposed based on the Bargmann-Michel-Telegdi equation for the spin dynamics \cite{Bargmann_1959}. An exemplary ab-initio numerical simulation based on the Dirac equation is carried out. Both, the intuitive explanation and the numerical simulation support the conclusion of the D-SFA on spin-effects. Unexpectedly,   magnitude and scaling of the spin-flip and spin-asymmetry effects at ionization are reduced dramatically, when the electron spin dynamics in the bound state is taken into account.

\begin{figure}
  \begin{center}
       \includegraphics[width=0.4\textwidth]{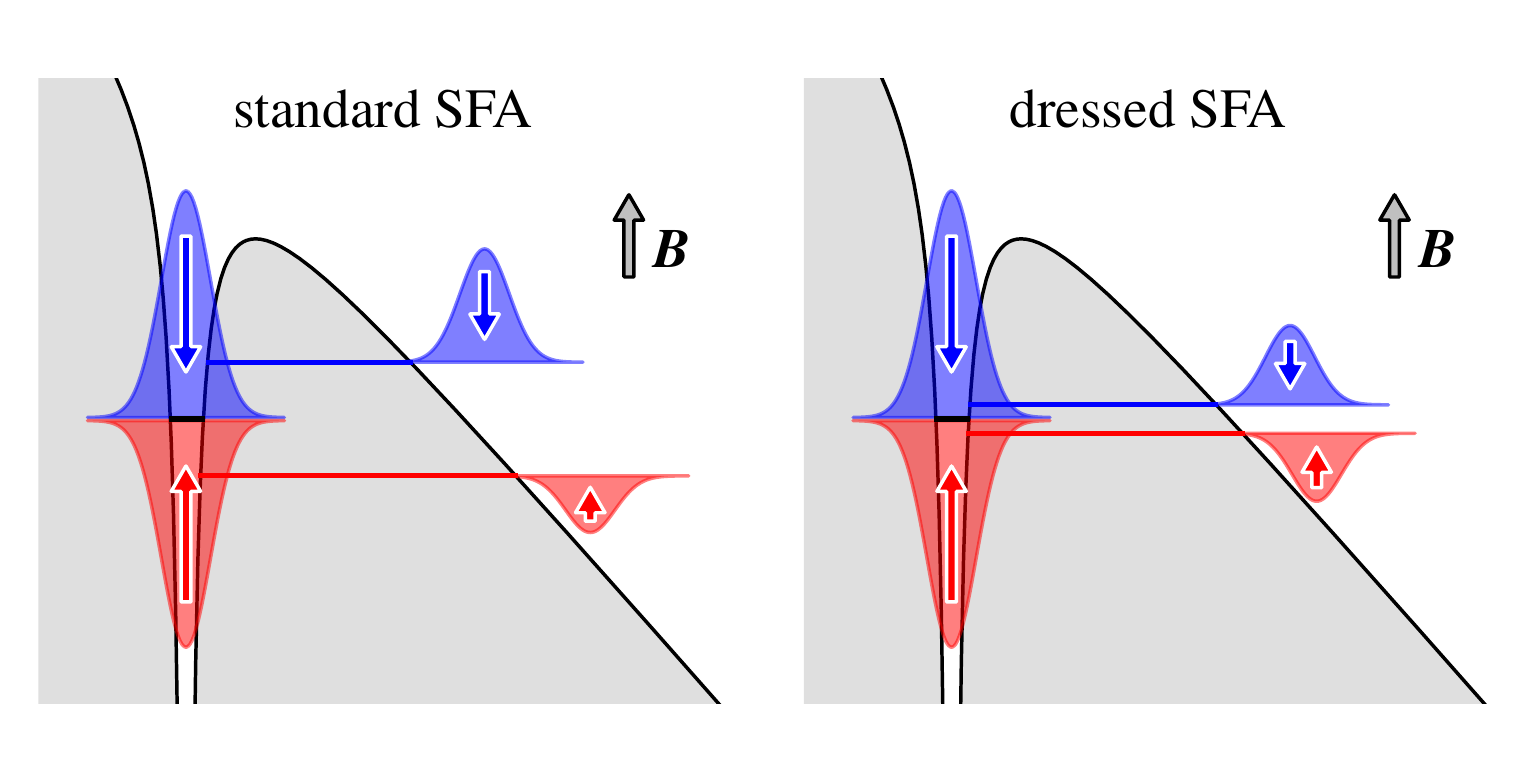}
 \includegraphics[width=0.4\textwidth]{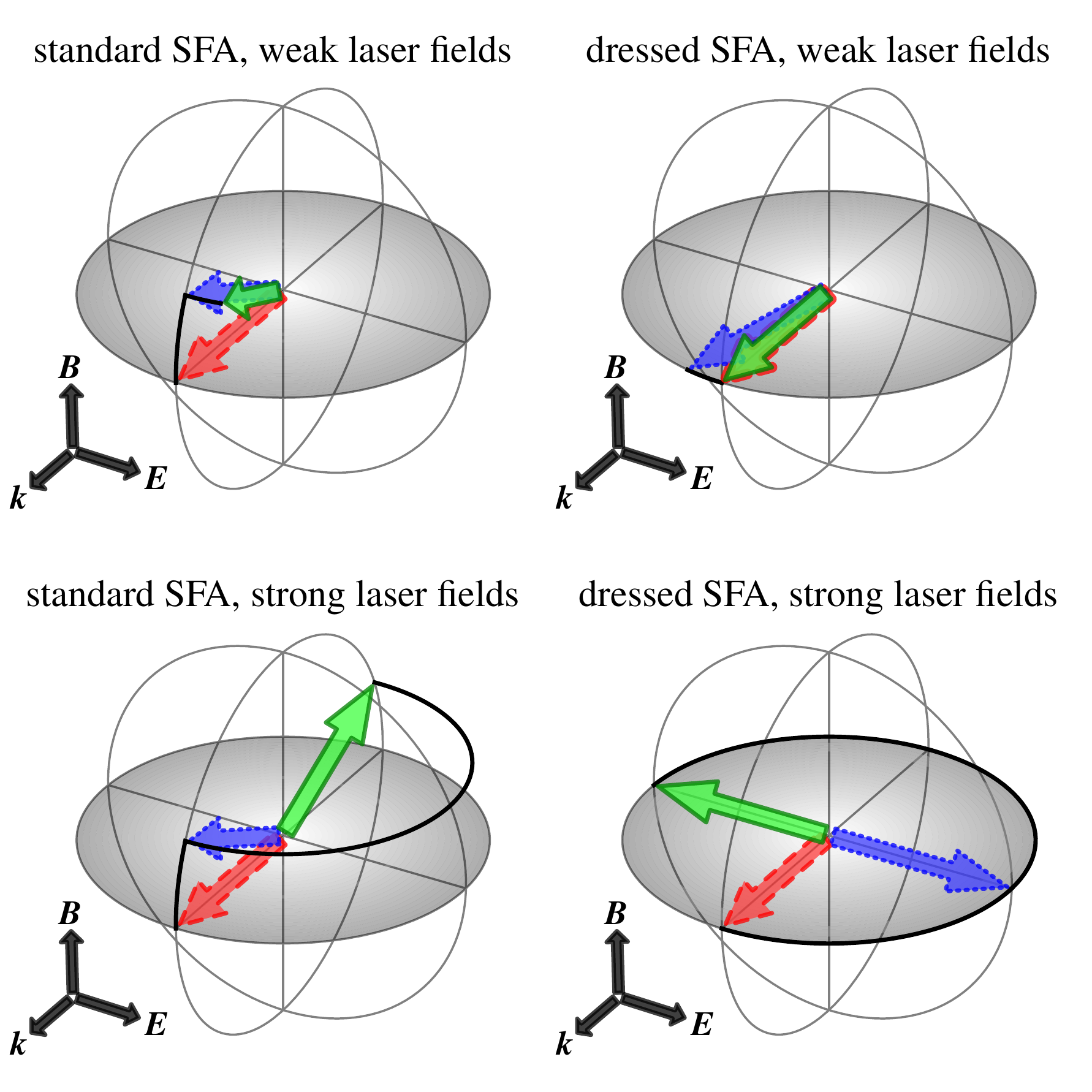}
        \caption{(Color online) Scheme illustrating the spin effects with the S-SFA (left column) and the D-SFA (right column). (top row) The initial spin is along the laser magnetic field: large asymmetry in the S-SFA and reduced asymmetry in the D-SFA. (middle row) Spin is along the laser propagation direction with weak fields $\mu \ll 1$, and (bottom row) with strong fields $\mu \gg 1$ ; (red, wiggled arrow), (blue, dotted arrow) and (green, solid arrow) indicate the initial spin, the spin after the tunneling, and the final spin, respectively. }
    \label{scheme}
  \end{center}
\end{figure}
\textit{SFA formalism.} Spin-resolved ionization rates are calculated employing a relativistic Coulomb-corrected SFA where the impact of  the Coulomb potential of the atomic core on the ionized electron continuum dynamics is treated via the eikonal approximation \cite{Gersten_1975,Avetissian_1997,Krainov_1997,Avetissian_1999,Smirnova_2008,Klaiber_2013a,Klaiber_2013b}. 
The differential ionization rate $dw/d^3\mathbf{p}=(\omega/2\pi)|M_{\rm f i}|^2$ is expressed via the SFA ionization transition matrix element  
\cite{Reiss_1990}
\begin{eqnarray}
  M_{\rm f i}=-i\int dt \,d^3 \mathbf{r}\,\psi_{\rm f}^\dagger(\mathbf{r},t)H_{\rm int}\phi^{(c)}_{\rm i}(\mathbf{r},t),
  \label{M_rel}
\end{eqnarray}
where $\phi^{(c)}_{\rm i}(\mathbf{r},t)$ and $\psi_{\rm f} (\mathbf{r},t)$ are the wave functions of the initial bound and the final continuum states, respectively, $H_{\rm int}$ is the interaction Hamiltonian, and  $\omega$ is the laser frequency (atomic units are used throughout). In the relativistic Coulomb-corrected SFA $\psi_{\rm f} (\mathbf{r},t)$ is the eikonal-Volkov wave function which is a solution of the Dirac equation for a continuum electron in the laser and Coulomb field in the eikonal approximation, see Eq. (42) in \cite{Klaiber_2013b}.
The interaction Hamiltonian $H_{\rm int}$ in Eq. (\ref{M_rel}) is defined via  a partitioning of the total Hamiltonian
\begin{eqnarray}
\hat{H}=c\boldsymbol{\alpha}\cdot(\hat{\textbf{p}}+\textbf{A})+\beta c^2+V^{(c)}-\Phi=\hat{H}_0+\hat{H}_{\rm int},\label{H}
\end{eqnarray}
where  the vector-potential of the laser field is chosen in the G{\"o}ppert-Mayer gauge
$A^{\mu}\equiv (\Phi,\mathbf{A})=(-\mathbf{r}\cdot\mathbf{E},-\hat{\mathbf{k}}\,(\mathbf{r}\cdot\mathbf{E})/c)$ \cite{Klaiber_2013a,Klaiber_2013b}, with the unit vector $\hat{\mathbf{k}}$ in the laser propagation direction. 
For the partition given by Eq. (\ref{H}), the initial state $ \phi^{(c)}_{\rm i}$ is governed by the equation \cite{Faisal_2007a}
\begin{eqnarray}
  i\partial_t \phi^{(c)}_{\rm i}=\hat{H}_0\phi^{(c)}_{\rm i}.
   \label{TSE}
\end{eqnarray}

\textit{Spin dynamics at tunneling and free motion in the continuum.} The S-SFA is based on the partition when $\hat{H}_0$ corresponds to the free atomic Hamiltonian: 
$\hat{H}_0^{(S)}=c\boldsymbol{\alpha}\cdot\hat{\textbf{p}}+\beta c^2+V^{(c)}$, 
$\hat{H}_{\rm int}^{(S)}=c\boldsymbol{\alpha}\cdot \textbf{A} -\Phi$,
and the spin dynamics in the bound state is completely neglected, i.e., before tunneling from the bound state  the electron spin is the same as in the initial state. In this case the spin effects are determined by the electron dynamics during tunneling and during the motion in the continuum. Because of the evident asymmetry in the spin evolution in this picture (frozen spin in the bound state and oscillating spin in the continuum) relatively large spin effects arise. On the other hand, it is well-known that the laser field can induce a large spin precession in the bound state \cite{Hu_1999,Hu_2001}. Moreover, the Zeeman-splitting of the bound state levels in the laser field can have impact on the tunneling probabilities, in this way inducing spin effects. Therefore, we use a D-SFA \cite{Klaiber_2013b} based on another partition of the Hamiltonian
\begin{eqnarray}
\hat{H}_0^{(D)}=c\boldsymbol{\alpha}\cdot [\hat{\mathbf{p}}-\hat{\mathbf{k}}(\mathbf{r}\cdot \mathbf{E})/c]+\beta c^2+V^{(c)}, \,\,\,\,\,\,\,
\hat{H}_{\rm int}^{(D)}=\mathbf{r}\cdot\mathbf{E},
\label{modified}
\end{eqnarray}
in which $\hat{H}_0^{(D)}$ includes the laser field, i.e., the bound state evolution in the laser field is accounted for, described by Eq.~(\ref{TSE}).

Let us begin with the case of a linearly polarized laser field. Spin effects in the tunneling regime of ionization are built up in three steps: spin precession in the bound state, spin rotation during tunneling, and spin precession during the electron free motion in the continuum, see Fig.~\ref{scheme}. Only the last two steps are included in the S-SFA. First, we  discuss the results of the S-SFA, to develop an intuitive picture for the last two steps of the spin evolution. Afterwards, we discuss how the picture changes when the first step is included within the D-SFA.

The S-SFA predicts a spin-asymmetry in ionization, i.e., the ionization probability depends on spin orientation with respect to the laser magnetic field 
\begin{eqnarray}
( w_{--} -w_{++})/w=2\sqrt{2I_p}/c,
     \label{w++--}
\end{eqnarray} 
where $w_{if}$ is the ionization probability from initial $i$ to final $f$ spin state, $\pm$ corresponds to the spin-up and -down state, respectively, $w=(w_{++} + w_{--}+w_{+-} + w_{-+})/2$ is the total ionization probability averaged over the possible initial spin states, and $I_p$ is the ionization potential. We assume that $E_0/E_a\ll 1$ (tunneling regime), and $I_p/c^2 \ll 1$ ($I_p/c^2\approx 0.25$ for hydrogenlike uranium), with the atomic field $E_a\equiv (2I_p)^{3/2}$ and the laser field amplitude $E_0$. 
The asymmetry (\ref{w++--}) has a simple explanation. In the S-SFA the laser quasi-static magnetic field is acting under the tunneling barrier, but not before tunneling. Since the total energy $-I_p$ is conserved before and during tunneling, the kinetic energy $\varepsilon_{\rm kin}$ under the barrier is shifted due to the spin magnetic field energy $\varepsilon_M=E_0/(2c)$ for the two spin orientations: $\varepsilon_{\rm kin}=-I_p\mp\varepsilon_M$. Since the kinetic energy determines the tunneling probability, the latter becomes spin-dependent: $w_{\pm}\propto \exp\left[- \frac{2^{5/2}}{3E_0}(I_p\pm\varepsilon_M)^{3/2}\right]\approx (1\mp \sqrt{2I_p}/c)e^{- \frac{2E_a}{3E_0}}$, which explains the asymmetry expressed by Eq. (\ref{w++--}) \footnote{We note that this
asymmetry only arises during one laser-half-cycle and will be compensated after a second consecutive half-cycle due to the
inverse magnetic field.}.

When the initial electron polarization is along the laser propagation direction, there is no spin-asymmetry but a spin-flip can occur
\begin{eqnarray}
   \frac{w_{+-}}{w}\approx  \frac{I_p}{2c^2}+ \frac{3}{8}\mu \,\, \left(\mu \ll 1\right);\,\,\,\,\frac{w_{+-}}{w}\approx 1- \frac{I_p}{2c^2}\,\, \left(\mu \gg 1\right)
  \label{flip_asym1}
\end{eqnarray}
where $\mu \equiv  (E_0/E_a)\xi_0^2$, $\xi_0\equiv E_0/(c\omega)$.
The $I_p/2c^2$-term in Eq. (\ref{flip_asym1}) originates from the tunneling step of ionization. 
The initial spin state polarized in the laser propagation direction $|\pm_{k}\rangle$ can be represented as a linear combination of spinors polarized in laser magnetic field direction $|\pm_{B}\rangle$. Because the tunneling probability depends on the spin projection along the magnetic field, see Eq. (\ref{w++--}), the coefficients of the superposition state will change during the tunneling, which is equivalent to the spin rotation in the $\textbf{B}$-$\hat{\mathbf{k}}$-plane:
$|\pm_{k}\rangle=\cos\frac{\theta}{2}e^{\mp i\frac{\pi}{4}}|\pm_{B}\rangle+\sin\frac{\theta}{2}e^{\pm i\frac{\pi}{4}}|\mp_{B}\rangle$, where $\theta$ is the angle of the spin with respect to the magnetic field, $\theta=\pi/2$ in the bound state. After tunneling, $\cos^2\frac{\theta}{2}=w_{++}/(w_{++}+w_{--})$ and $\sin^2\frac{\theta}{2}=w_{--}/(w_{++}+w_{--})$, i.e., $\cos\theta  \approx \sqrt{2I_p}/c$. Thus, tunneling induces a spin rotation by the angle $\Delta\theta=\pi/2-\theta\approx  \sqrt{2I_p}/c$. From $\langle s_k\rangle=\cos\Delta\theta$ and $w_{+-}/w=(1-\langle s_k\rangle)/2$, the $I_p/2c^2$-term in Eq. (\ref{flip_asym1}) is derived.

The terms in the spin-flip besides of $I_p/2c^2$ in Eq.~(\ref{flip_asym1}) are due to the third step: the electron motion in the laser field after exiting the tunneling barrier, which can be described by a semi-classical Bargmann-Michel-Telegdi equation \cite{Bargmann_1959}, attributing to the classical electron a spin angular momentum $\textbf{s}$,
\begin{eqnarray}
 \frac{d\textbf{s}}{dt}=\frac{1}{\gamma} \textbf{s}\times  (\hat{\textbf{k}}\times \textbf{E})- \frac{1}{\gamma+1} \textbf{s}\times (\boldsymbol{\beta}\times \textbf{E}),
\end{eqnarray}
with $\boldsymbol{\beta}$- and $\gamma$-Lorentz factors. Using the classical equations of motions in a laser field and introducing the precession angle $\varphi$ in the $\hat{\textbf{k}}$-$\textbf{E}$-plane via $s_k=s\cos\varphi$ and $s_E =s\sin\varphi$, where indices $k$ and $E$ indicate the vector components along the laser propagation and polarization direction, respectively, the spin precession equation is derived: $\frac{d\varphi}{d\eta}=\Omega(\eta)$,
with $\Omega(\eta)=-\xi (\eta) \cos\eta/\{1+[\xi (\eta)\sin\eta-\xi_0\sin\eta_0]^2/4\}$, where $\eta_0$ is the ionization phase, $\xi (\eta)=E(\eta)/c\omega$, and $E(\eta)$ the envelope of the laser field. The denominator in the expression for $\Omega(\eta)$  originates from the Lorentz-boost factor $1- p_k/[c(1+\gamma)]$.
After switching off the laser field, the spin precession angle is 
$ \Delta\varphi\equiv \varphi-\varphi_0=2\arctan\left(\frac{\xi_0}{2}\sin\eta_0\right)=-2\arctan\left(\frac{p_E}{2c}\right)$,
where $\varphi=\varphi_0$ at the tunneling exit, and $p_E=-c\xi_0\sin\eta_0$ is the final electron momentum along the laser polarization direction. Averaging over $p_E$ yields at $\mu\ll 1$: $\Delta\varphi^2\approx \Delta p_E^2/(2c^2)=(3/2)\mu^2$,  with the momentum distribution width $\Delta p_E$ \cite{Popov_2004u}.
The electron spin-state after the rotation around the magnetic field axis by an angle $\varphi(\eta)$ due to the electron free motion in the laser field, is $|\pm_{k}\rangle=\cos\frac{\theta}{2}e^{i\frac{\varphi(\eta)}{2}\mp i\frac{\pi}{4}}|\pm_{B}\rangle+\sin\frac{\theta}{2}e^{-i\frac{\varphi(\eta)}{2}\pm i\frac{\pi}{4}}|\mp_{B}\rangle$. After the interaction $\langle s_k\rangle=\sin\theta\cos\varphi =\cos\Delta\theta \cos\Delta\varphi\approx 1- \Delta\theta^2/2 - \Delta\varphi^2/2$,
with $\Delta\theta^2=2I_p/c^2$ and $\Delta\varphi^2=(3/2)\mu^2$,
which leads to Eq.~(\ref{flip_asym1}) for the spin-flip probability at $\mu\ll 1$.
Thus, when the electron appears in the continuum nonadiabatically in a certain laser phase, the symmetry of the spin precession in the laser field is broken, resulting in an effective spin-flip at switching off the laser field. 
In strong laser fields with $\mu\gg 1$, the spin precession frequency due to free motion in the laser field initially is large $\sim \xi_0$ but quickly vanishes due to the Lorentz-boost factor in $\Omega(\eta)$, resulting to $\varphi\approx\pi$, i.e., to a complete spin-flip  $w_{+-}/w\approx 1$, which corresponds to the first term in Eq. (\ref{flip_asym1}) at $\mu\gg 1$.

\begin{figure}
  \begin{center}
       \includegraphics[width=0.239\textwidth]{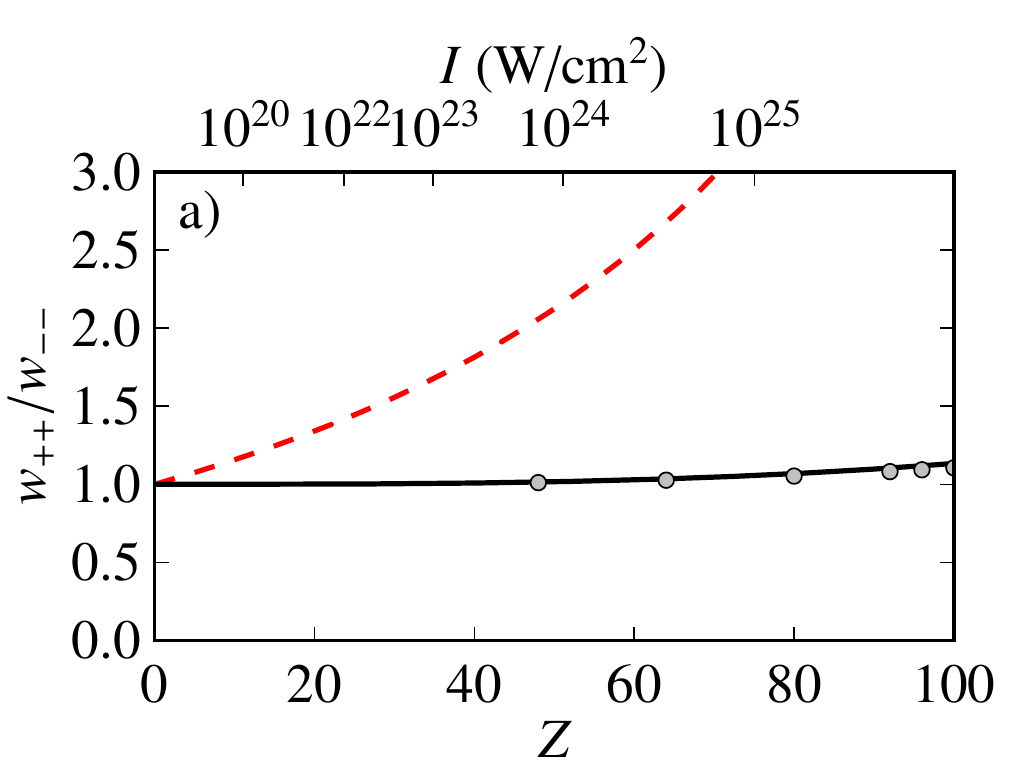}
\includegraphics[width=0.239\textwidth]{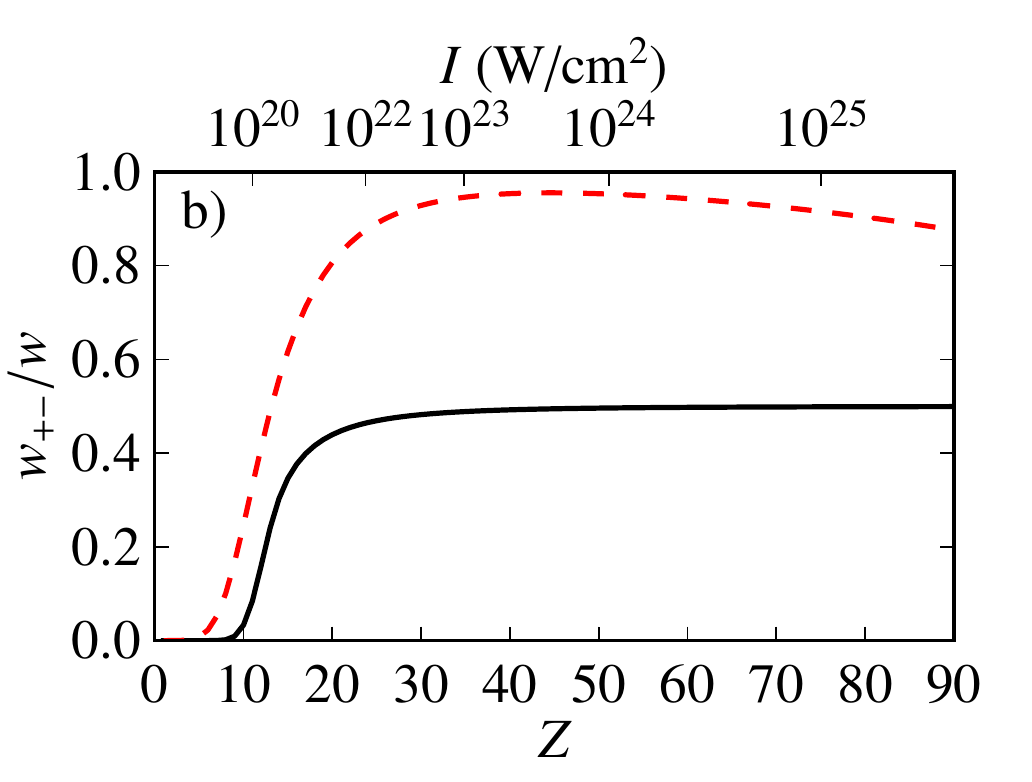}
        \caption{(Color online) (a) spin-asymmetry factor $w_{++}/w_{--}$ when the electron initial spin is along the laser magnetic field; (b) Spin-flip probability when the initial spin is along the laser propagation direction; (black, solid) with D-SFA, (red, dashed) with S-SFA, ab initio numerical calculations are shown with circles; $\omega=0.05$ a.u. and $E_0/E_a=1/25$.}
    \label{asym}
  \end{center}
\end{figure}

\textit{Spin-asymmetry.} Now let us discuss how spin effects are modified when one takes into account the spin precession in the bound state and the Zeeman-splitting of the bound energy levels in the laser field. For this purpose we calculate the spin resolved ionization probabilities in the D-SFA based on the partition of Eq.~(\ref{modified}). When the initial spin is aligned along the laser magnetic field, there occurs no spin precession neither in the bound, nor in the free electron state. However, in contrast to the S-SFA, here the energy of the bound state is shifted due to Zeeman-splitting.
As a consequence, the spin-asymmetry in ionization, which existed in the S-SFA, is significantly reduced [see also Fig.~\ref{asym}~(a)]:
\begin{eqnarray}
  w_{++}/w_{--}=\mathfrak{W}
 e^{-\frac{2 \Xi}{\sqrt{3}}  \left(\frac{1}{\sqrt{\Xi ^2+1}}+\frac{4}{\Xi ^2+1}-2\right)}
   \approx 1+\left(2I_p/c^2\right)^{3/2}
\end{eqnarray}
with $\Xi=\sqrt{1-\Upsilon/2(\sqrt{\Upsilon^2+8}-\Upsilon)}$, $\Upsilon=1-I_p/c^2$, and $\mathfrak{W}=\left(7 \Xi ^2+4 \sqrt{3} \sqrt{\Xi ^2+1} \Xi \right)/\left(\sqrt{3}-\Xi \right)^2$.
The reduced asymmetry can be understood as follows. Due to the electron spin interaction with the laser magnetic field, the electron has an additional energy ($\sim B_0/c$). However, this energy is in leading order the same as in the bound as well as in the free state.   The spin energy adds to or subtracts from the total electron energy for the spin-up or spin-down cases, respectively, but the kinetic energy due to the motion along the laser field that is crucial for the tunneling is unaffected and is the same for any spin value. Therefore, the ionization probability does not depend on the spin,
i.e., there is no spin-asymmetry in ionization in leading order by $I_p/c^2$. In higher orders a small asymmetry remains of the order of $(\sqrt{2I_p/c^2})^3$.    
It is due to the difference in the rest-frame magnetic field in the bound and Volkov states. In the electron rest-frame there is an additional magnetic field $B^{\prime}\approx \boldsymbol{\beta}\times \mathbf{E}\sim p_k E_0/c$. The momentum difference in the bound and the free state at the tunneling can be estimated $p_k\sim I_p/c$~\cite{Klaiber_2013c}, which will lead to an asymmetry $(w_{++}-w_{--})/w\sim \sqrt{I_p/c^2}B^{\prime}/B\sim (I_p/c^2)^{3/2}$. Thus, due to Zeeman splitting of the bound state energy, the spin-asymmetry is of order $(I_p/c^2)^{3/2}$, significantly less than the prediction of the S-SFA. For example, for Ne$^{9+}$ the asymmetry factor is only 1.0004 in the D-SFA, while it is 1.15 in the S-SFA. To confirm these results, we have carried out ab initio numerical calculations of the ionization probability at different initial spin states via Dirac equation. The ionization rate is deduced from the bound state depletion rate.  Fig~\ref{asym}~(a) 
shows that the  numerical result can be well reproduced by D-SFA. 

Note that the spin-asymmetry for the differential probabilities can be significantly larger than for the total one. In fact, the spin-asymmetry parameter on the wings of the momentum distribution can be estimated to be larger than the total one by a factor $\sqrt{(E_0/E_a)(c^2/I_p)}$ which can amount to an order of magnitude.

\textit{Spin-flip.} We consider, using D-SFA, the role of the spin precession in the bound state on the spin-flip effect at ionization. 
Spin-flip happens when the initial polarization is along the laser propagation direction. In solving Eq.~(\ref{TSE}) for the electron bound state, we restrict the expansion basis of the time-dependent wave function of the bound state only over the spin quantum number, representing it as a linear combination of the states with the initial spin along the laser magnetic field:
$ \tilde{\phi}^c_{k,\pm}(t)=\frac{1}{\sqrt{2}} \left[\tilde{\phi}^c_{B,\pm}(t)-i\tilde{\phi}^c_{B,\mp}(t)\right]$ \cite{Klaiber_2013b}.
Here $\tilde{\phi}^{(c)}_{B,\pm}(t)$ is the time-dependent wave function of the ground state of a hydrogenlike ion that has an initial spin along the laser magnetic field direction: 
$\tilde{\phi}^{(c)}_{B,\pm}(t)=\phi^{(c)}_{B,\pm}(t)\exp\left[\pm i\frac{\tilde{A}(\eta)}{2c}\left(1-\frac{2I_p}{3c^2}\right)\right]$. These states have been used in the calculation of the spin-dependent ionization probabilities via Eq. (\ref{M_rel}). 
The results of the D-SFA calculations are summarized in Fig.~\ref{asym}~(b). The behavior of the spin-flip probability is quite different from the S-SFA result. In weak laser fields $\mu\ll 1$, there occurs almost no spin-flip:
\begin{eqnarray}
w_{+-}/w \approx {\mathcal O} [(I_p/c^2)^3]+{\mathcal O}[(I_p/c^2)^2\mu],
  \end{eqnarray}
which is in contrast to the large spin-flip prediction in the S-SFA, see Eq. (\ref{flip_asym1}). Whereas in strong fields $\mu\gg 1$,
the probabilities for the spin-up and -down states are almost equal in the final state:
\begin{eqnarray}
w_{+-}/w \approx 1/2+{\mathcal O}(1/\mu),
  \end{eqnarray}
while the S-SFA gives almost total spin-flip in this limit.

The modification of the spin-flip probability when the spin precession in the bound state is taken into account (D-SFA), has a simple intuitive explanation.
In the bound state a precession of the electron spin occurs with a frequency $\Omega_b\approx \omega\xi_0(1-2I_p/3c^2)\sim \omega\xi_0$ (cf. \cite{Hu_1999}). For the electron in continuum, the spin precession frequency is $\Omega(\eta)\sim \omega\xi_0$ in weak fields $\mu\ll 1$ (the Lorentz-boost-factor has only a little influence) which coincides with the spin dynamics in the bound state. Consequently, there is no asymmetry in the spin dynamics due to tunneling and the spin state does not change when adiabatically switching on and off the laser field. In opposite, when the laser field is strong $\mu\gg 1$, the bound and free spin dynamics are different. The electron spin in the bound state still exhibits rapid precession with $\Omega_b$. During the dominant ionization phase $\delta\eta_i$, the spin has many different orientations because of $\Omega_b\delta \eta_i /\omega\gg 1$, with $\delta\eta_i \sim \sqrt{E_0/E_a}$ \footnote{The dominant phase interval for ionization $\delta\eta_i$ can be estimated from 
the tunneling exponent: $\exp[-E_a/(E_0\cos \eta)]\approx \exp[-E_a/E_0-E_a\delta\eta^2/(2E_0)]$, from the latter the typical ionization phase is $\delta\eta_i \sim \sqrt{E_0/E_a}$.}, and on average spin-up and -down positions are equally probable, i.e., the spin-flip probability is approximately equal to 1/2. After the electron has tunneled into continuum, the electron drifts into the laser propagation direction and feels a slowly varying laser phase due to the Lorentz-boost factor. During the continuum motion in the laser pulse,  a total  spin-flip occurs  $\varphi\approx\pi$ at $\mu\gg 1$. However,
the probability ratio of the spin-up and -down states remain equally probable at the end of the pulse.

\textit{Circular polarization.} Spin effects in a circularly polarized laser field are qualitatively the same as in the case of linear polarization. According to D-SFA, there is no spin-asymmetry in ionization. In a weak field, $\xi_0\ll 1$, the spin-flip probability is vanishing $w_{+-}\approx 0$ (the spin quantization axis is chosen along the laser propagation direction), while in a strong field $\xi_0\gg 1$, it is $w_{+-}/w\approx 1/2$, which have the same explanation as in the case of a linearly polarized field.

\textit{Experimental observability.} Although, our calculations show that the spin-asymmetry is rather small and difficult for observation, the spin-flip can have rather large values in the strong field limit. For typical experimental parameters, e.g., ionization of hydrogenlike Ne$^{9+}$-ions in a strong infrared laser field with an intensity of $8.5\times 10^{19}$ W/cm$^{2}$, the D-SFA predicts a relative spin-flip probability of about 0.1. Employing Fe$^{25+}$-ions  and a laser field with an intensity of $I=1.7\times 10^{22}$ W/cm$^2$, a spin-flip probability of approximately 0.4 can be achieved. These measurements require an initially polarized target of ions.

The electron spin flip can also be revealed via measurement of ion parameters. The angular momentum change of the ion during ionization $\Delta J_i$ can be related to the electron spin change $\Delta S$, using the angular momentum and energy conservation laws: $n=\Delta J_i+\Delta J_e$ and $n\omega=I_p+2U_p$ in the case of circular polarization, where $\Delta J_e=\Delta L_e+\Delta S$, $\Delta L_e=(2U_p+I_p)/\omega$ is the electron orbital angular momentum change \cite{Reiss_1990} and $n$ the number of absorbed photons. Then, $\Delta S=-\Delta J_i$ could be measured via the magnetic moment of the ionic core.

\textit{Conclusion.} In the relativistic regime of tunneling ionization with super-strong laser fields large spin-flip effects are measurable when employing highly charged ions, initially polarized along the laser propagation direction. The spin-asymmetry in the ionization is rather small. Our ab initio numerical calculations confirms the physical relevance of the results of D-SFA. Our study shows that, even if an electron is very tightly bound to an ionic core, it may still be crucially affected by a laser field of moderate intensity.

We acknowledge valuable discussions with C. H. Keitel.

\bibliography{strong_fields_bibliography}

\begin{thebibliography}{48}
\expandafter\ifx\csname natexlab\endcsname\relax\def\natexlab#1{#1}\fi
\expandafter\ifx\csname bibnamefont\endcsname\relax
  \def\bibnamefont#1{#1}\fi
\expandafter\ifx\csname bibfnamefont\endcsname\relax
  \def\bibfnamefont#1{#1}\fi
\expandafter\ifx\csname citenamefont\endcsname\relax
  \def\citenamefont#1{#1}\fi
\expandafter\ifx\csname url\endcsname\relax
  \def\url#1{\texttt{#1}}\fi
\expandafter\ifx\csname urlprefix\endcsname\relax\def\urlprefix{URL }\fi
\providecommand{\bibinfo}[2]{#2}
\providecommand{\eprint}[2][]{\url{#2}}

\bibitem[{\citenamefont{Piazza et~al.}(2012)\citenamefont{Piazza, M\"uller,
  Hatsagortsyan, and Keitel}}]{RMP_2012}
\bibinfo{author}{\bibfnamefont{A.~D.} \bibnamefont{Piazza}},
  \bibinfo{author}{\bibfnamefont{C.}~\bibnamefont{M\"uller}},
  \bibinfo{author}{\bibfnamefont{K.~Z.} \bibnamefont{Hatsagortsyan}},
  \bibnamefont{and} \bibinfo{author}{\bibfnamefont{C.~H.}
  \bibnamefont{Keitel}}, \bibinfo{journal}{Rev. Mod. Phys.}
  \textbf{\bibinfo{volume}{84}}, \bibinfo{pages}{1177} (\bibinfo{year}{2012}).

\bibitem[{\citenamefont{Moore et~al.}(1999)\citenamefont{Moore, Ting, McNaught,
  Qiu, Burris, and Sprangle}}]{Moore_1999}
\bibinfo{author}{\bibfnamefont{C.~I.} \bibnamefont{Moore}},
  \bibinfo{author}{\bibfnamefont{A.}~\bibnamefont{Ting}},
  \bibinfo{author}{\bibfnamefont{S.~J.} \bibnamefont{McNaught}},
  \bibinfo{author}{\bibfnamefont{J.}~\bibnamefont{Qiu}},
  \bibinfo{author}{\bibfnamefont{H.~R.} \bibnamefont{Burris}},
  \bibnamefont{and} \bibinfo{author}{\bibfnamefont{P.}~\bibnamefont{Sprangle}},
  \bibinfo{journal}{Phys. Rev. Lett.} \textbf{\bibinfo{volume}{82}},
  \bibinfo{pages}{1688} (\bibinfo{year}{1999}).

\bibitem[{\citenamefont{Chowdhury et~al.}(2001)\citenamefont{Chowdhury, Barty,
  and Walker}}]{Chowdhury_2001}
\bibinfo{author}{\bibfnamefont{E.~A.} \bibnamefont{Chowdhury}},
  \bibinfo{author}{\bibfnamefont{C.~P.~J.} \bibnamefont{Barty}},
  \bibnamefont{and} \bibinfo{author}{\bibfnamefont{B.~C.}
  \bibnamefont{Walker}}, \bibinfo{journal}{Phys. Rev. A}
  \textbf{\bibinfo{volume}{63}}, \bibinfo{pages}{042712}
  (\bibinfo{year}{2001}).

\bibitem[{\citenamefont{Dammasch et~al.}(2001)\citenamefont{Dammasch, D\"orr,
  Eichmann, Lenz, and Sandner}}]{Dammasch_2001}
\bibinfo{author}{\bibfnamefont{M.}~\bibnamefont{Dammasch}},
  \bibinfo{author}{\bibfnamefont{M.}~\bibnamefont{D\"orr}},
  \bibinfo{author}{\bibfnamefont{U.}~\bibnamefont{Eichmann}},
  \bibinfo{author}{\bibfnamefont{E.}~\bibnamefont{Lenz}}, \bibnamefont{and}
  \bibinfo{author}{\bibfnamefont{W.}~\bibnamefont{Sandner}},
  \bibinfo{journal}{Phys. Rev. A} \textbf{\bibinfo{volume}{64}},
  \bibinfo{pages}{061402} (\bibinfo{year}{2001}).

\bibitem[{\citenamefont{Yamakawa et~al.}(2003)\citenamefont{Yamakawa, Akahane,
  Fukuda, Aoyama, Inoue, and Ueda}}]{Yamakawa_2003}
\bibinfo{author}{\bibfnamefont{K.}~\bibnamefont{Yamakawa}},
  \bibinfo{author}{\bibfnamefont{Y.}~\bibnamefont{Akahane}},
  \bibinfo{author}{\bibfnamefont{Y.}~\bibnamefont{Fukuda}},
  \bibinfo{author}{\bibfnamefont{M.}~\bibnamefont{Aoyama}},
  \bibinfo{author}{\bibfnamefont{N.}~\bibnamefont{Inoue}}, \bibnamefont{and}
  \bibinfo{author}{\bibfnamefont{H.}~\bibnamefont{Ueda}},
  \bibinfo{journal}{Phys. Rev. A} \textbf{\bibinfo{volume}{68}},
  \bibinfo{pages}{065403} (\bibinfo{year}{2003}).

\bibitem[{\citenamefont{Gubbini}(2005)}]{Gubbini_2005}
\bibinfo{author}{\bibfnamefont{E.}~\bibnamefont{Gubbini}}, \bibinfo{journal}{J.
  Phys. B} \textbf{\bibinfo{volume}{38}}, \bibinfo{pages}{L87}
  (\bibinfo{year}{2005}).

\bibitem[{\citenamefont{DiChiara et~al.}(2008)\citenamefont{DiChiara,
  Ghebregziabher, Sauer, Waesche, Palaniyappan, Wen, and
  Walker}}]{DiChiara_2008}
\bibinfo{author}{\bibfnamefont{A.~D.} \bibnamefont{DiChiara}},
  \bibinfo{author}{\bibfnamefont{I.}~\bibnamefont{Ghebregziabher}},
  \bibinfo{author}{\bibfnamefont{R.}~\bibnamefont{Sauer}},
  \bibinfo{author}{\bibfnamefont{J.}~\bibnamefont{Waesche}},
  \bibinfo{author}{\bibfnamefont{S.}~\bibnamefont{Palaniyappan}},
  \bibinfo{author}{\bibfnamefont{B.~L.} \bibnamefont{Wen}}, \bibnamefont{and}
  \bibinfo{author}{\bibfnamefont{B.~C.} \bibnamefont{Walker}},
  \bibinfo{journal}{Phys. Rev. Lett.} \textbf{\bibinfo{volume}{101}},
  \bibinfo{pages}{173002} (\bibinfo{year}{2008}).

\bibitem[{\citenamefont{Palaniyappan et~al.}(2008)\citenamefont{Palaniyappan,
  Mitchell, Sauer, Ghebregziabher, White, Decamp, and
  Walker}}]{Palaniyappan_2008}
\bibinfo{author}{\bibfnamefont{S.}~\bibnamefont{Palaniyappan}},
  \bibinfo{author}{\bibfnamefont{R.}~\bibnamefont{Mitchell}},
  \bibinfo{author}{\bibfnamefont{R.}~\bibnamefont{Sauer}},
  \bibinfo{author}{\bibfnamefont{I.}~\bibnamefont{Ghebregziabher}},
  \bibinfo{author}{\bibfnamefont{S.~L.} \bibnamefont{White}},
  \bibinfo{author}{\bibfnamefont{M.~F.} \bibnamefont{Decamp}},
  \bibnamefont{and} \bibinfo{author}{\bibfnamefont{B.~C.}
  \bibnamefont{Walker}}, \bibinfo{journal}{Phys. Rev. Lett.}
  \textbf{\bibinfo{volume}{100}}, \bibinfo{pages}{183001}
  (\bibinfo{year}{2008}).

\bibitem[{\citenamefont{DiChiara et~al.}(2010)\citenamefont{DiChiara,
  Ghebregziabher, Waesche, Stanev, Ekanayake, Barclay, Wells, Watts, Videtto,
  Mancuso et~al.}}]{DiChiara_2010}
\bibinfo{author}{\bibfnamefont{A.~D.} \bibnamefont{DiChiara}},
  \bibinfo{author}{\bibfnamefont{I.}~\bibnamefont{Ghebregziabher}},
  \bibinfo{author}{\bibfnamefont{J.~M.} \bibnamefont{Waesche}},
  \bibinfo{author}{\bibfnamefont{T.}~\bibnamefont{Stanev}},
  \bibinfo{author}{\bibfnamefont{N.}~\bibnamefont{Ekanayake}},
  \bibinfo{author}{\bibfnamefont{L.~R.} \bibnamefont{Barclay}},
  \bibinfo{author}{\bibfnamefont{S.~J.} \bibnamefont{Wells}},
  \bibinfo{author}{\bibfnamefont{A.}~\bibnamefont{Watts}},
  \bibinfo{author}{\bibfnamefont{M.}~\bibnamefont{Videtto}},
  \bibinfo{author}{\bibfnamefont{C.~A.} \bibnamefont{Mancuso}},
  \bibnamefont{et~al.}, \bibinfo{journal}{Phys. Rev. A}
  \textbf{\bibinfo{volume}{81}}, \bibinfo{pages}{043417}
  (\bibinfo{year}{2010}).

\bibitem[{\citenamefont{{The Petawatt High-Energy Laser for heavy Ion
  eXperiments (PHELIX) facility}}(2011)}]{PHELIX}
\bibinfo{author}{\bibnamefont{{The Petawatt High-Energy Laser for heavy Ion
  eXperiments (PHELIX) facility}}} (\bibinfo{year}{2011}),
  \urlprefix\url{http://www.gsi.de/forschung/pp/phelix/index_e.html}.

\bibitem[{\citenamefont{Reiss}(1990{\natexlab{a}})}]{Reiss_1990}
\bibinfo{author}{\bibfnamefont{H.~R.} \bibnamefont{Reiss}},
  \bibinfo{journal}{Phys. Rev. A} \textbf{\bibinfo{volume}{42}},
  \bibinfo{pages}{1476} (\bibinfo{year}{1990}{\natexlab{a}}).

\bibitem[{\citenamefont{Reiss}(1990{\natexlab{b}})}]{Reiss_1990b}
\bibinfo{author}{\bibfnamefont{H.~R.} \bibnamefont{Reiss}},
  \bibinfo{journal}{J. Opt. Soc. Am. B} \textbf{\bibinfo{volume}{7}},
  \bibinfo{pages}{574} (\bibinfo{year}{1990}{\natexlab{b}}).

\bibitem[{\citenamefont{Popov et~al.}(1997)\citenamefont{Popov, Mur, and
  Karnakov}}]{Popov_1997}
\bibinfo{author}{\bibfnamefont{V.}~\bibnamefont{Popov}},
  \bibinfo{author}{\bibfnamefont{V.}~\bibnamefont{Mur}}, \bibnamefont{and}
  \bibinfo{author}{\bibfnamefont{B.}~\bibnamefont{Karnakov}},
  \bibinfo{journal}{JETP Letters} \textbf{\bibinfo{volume}{66}},
  \bibinfo{pages}{229} (\bibinfo{year}{1997}).

\bibitem[{\citenamefont{Mur et~al.}(1998)\citenamefont{Mur, Karnakov, and
  Popov}}]{Mur_1998}
\bibinfo{author}{\bibfnamefont{V.}~\bibnamefont{Mur}},
  \bibinfo{author}{\bibfnamefont{B.}~\bibnamefont{Karnakov}}, \bibnamefont{and}
  \bibinfo{author}{\bibfnamefont{V.}~\bibnamefont{Popov}},
  \bibinfo{journal}{JETP Letters} \textbf{\bibinfo{volume}{87}},
  \bibinfo{pages}{433} (\bibinfo{year}{1998}).

\bibitem[{\citenamefont{Popov et~al.}(2004)\citenamefont{Popov, Karnakov, and
  Mur}}]{Popov_2004}
\bibinfo{author}{\bibfnamefont{V.~S.} \bibnamefont{Popov}},
  \bibinfo{author}{\bibfnamefont{B.~M.} \bibnamefont{Karnakov}},
  \bibnamefont{and} \bibinfo{author}{\bibfnamefont{V.~D.} \bibnamefont{Mur}},
  \bibinfo{journal}{Zh. Exp. Theor. Fiz.} \textbf{\bibinfo{volume}{79}},
  \bibinfo{pages}{320} (\bibinfo{year}{2004}).

\bibitem[{\citenamefont{Milosevic
  et~al.}(2002{\natexlab{a}})\citenamefont{Milosevic, Krainov, and
  Brabec}}]{Milosevic_2002r1}
\bibinfo{author}{\bibfnamefont{N.}~\bibnamefont{Milosevic}},
  \bibinfo{author}{\bibfnamefont{V.~P.} \bibnamefont{Krainov}},
  \bibnamefont{and} \bibinfo{author}{\bibfnamefont{T.}~\bibnamefont{Brabec}},
  \bibinfo{journal}{Phys. Rev. Lett.} \textbf{\bibinfo{volume}{89}},
  \bibinfo{pages}{193001} (\bibinfo{year}{2002}{\natexlab{a}}).

\bibitem[{\citenamefont{Milosevic
  et~al.}(2002{\natexlab{b}})\citenamefont{Milosevic, Krainov, and
  Brabec}}]{Milosevic_2002r2}
\bibinfo{author}{\bibfnamefont{N.}~\bibnamefont{Milosevic}},
  \bibinfo{author}{\bibfnamefont{V.~P.} \bibnamefont{Krainov}},
  \bibnamefont{and} \bibinfo{author}{\bibfnamefont{T.}~\bibnamefont{Brabec}},
  \bibinfo{journal}{J. Phys. B} \textbf{\bibinfo{volume}{35}},
  \bibinfo{pages}{3515} (\bibinfo{year}{2002}{\natexlab{b}}).

\bibitem[{\citenamefont{Becker et~al.}(2002)\citenamefont{Becker, Grasbon,
  Kopold, Milo\u{s}evi\'c, Paulus, and Walther}}]{Becker_2002}
\bibinfo{author}{\bibfnamefont{W.}~\bibnamefont{Becker}},
  \bibinfo{author}{\bibfnamefont{F.}~\bibnamefont{Grasbon}},
  \bibinfo{author}{\bibfnamefont{R.}~\bibnamefont{Kopold}},
  \bibinfo{author}{\bibfnamefont{D.}~\bibnamefont{Milo\u{s}evi\'c}},
  \bibinfo{author}{\bibfnamefont{G.~G.} \bibnamefont{Paulus}},
  \bibnamefont{and} \bibinfo{author}{\bibfnamefont{H.}~\bibnamefont{Walther}},
  \bibinfo{journal}{Adv. Atom. Mol. Opt. Phys.} \textbf{\bibinfo{volume}{48}},
  \bibinfo{pages}{35} (\bibinfo{year}{2002}).

\bibitem[{\citenamefont{Klaiber et~al.}(2005)\citenamefont{Klaiber,
  Hatsagortsyan, and Keitel}}]{Klaiber_2005}
\bibinfo{author}{\bibfnamefont{M.}~\bibnamefont{Klaiber}},
  \bibinfo{author}{\bibfnamefont{K.~Z.} \bibnamefont{Hatsagortsyan}},
  \bibnamefont{and} \bibinfo{author}{\bibfnamefont{C.~H.}
  \bibnamefont{Keitel}}, \bibinfo{journal}{Phys. Rev. A}
  \textbf{\bibinfo{volume}{71}}, \bibinfo{pages}{033408}
  (\bibinfo{year}{2005}).

\bibitem[{\citenamefont{Faisal and Bhattacharyya}(2004)}]{Faisal_2004}
\bibinfo{author}{\bibfnamefont{F.~H.~M.} \bibnamefont{Faisal}}
  \bibnamefont{and}
  \bibinfo{author}{\bibfnamefont{S.}~\bibnamefont{Bhattacharyya}},
  \bibinfo{journal}{Phys. Rev. Lett.} \textbf{\bibinfo{volume}{93}},
  \bibinfo{pages}{053002} (\bibinfo{year}{2004}).

\bibitem[{\citenamefont{Bunkin et~al.}(1972)\citenamefont{Bunkin, Kazakov, and
  Fedorov}}]{Bunkin_1972}
\bibinfo{author}{\bibfnamefont{F.}~\bibnamefont{Bunkin}},
  \bibinfo{author}{\bibfnamefont{A.}~\bibnamefont{Kazakov}}, \bibnamefont{and}
  \bibinfo{author}{\bibfnamefont{M.}~\bibnamefont{Fedorov}},
  \bibinfo{journal}{Usp. Fiz. Nauk} \textbf{\bibinfo{volume}{107}},
  \bibinfo{pages}{559} (\bibinfo{year}{1972}), \bibinfo{note}{[Sov. Phys. Usp.
  \textbf{15}, 416 (1973)]}.

\bibitem[{\citenamefont{Szymanowski et~al.}(1997)\citenamefont{Szymanowski,
  V\'{e}niard, Ta\"{i}eb, Maquet, and Keitel}}]{Szymanowski_1997}
\bibinfo{author}{\bibfnamefont{C.}~\bibnamefont{Szymanowski}},
  \bibinfo{author}{\bibfnamefont{V.}~\bibnamefont{V\'{e}niard}},
  \bibinfo{author}{\bibfnamefont{R.}~\bibnamefont{Ta\"{i}eb}},
  \bibinfo{author}{\bibfnamefont{A.}~\bibnamefont{Maquet}}, \bibnamefont{and}
  \bibinfo{author}{\bibfnamefont{C.~H.} \bibnamefont{Keitel}},
  \bibinfo{journal}{Phys. Rev. A} \textbf{\bibinfo{volume}{56}},
  \bibinfo{pages}{3846} (\bibinfo{year}{1997}).

\bibitem[{\citenamefont{Walser et~al.}(1999)\citenamefont{Walser, Szymanowski,
  and Keitel}}]{Walser_1999}
\bibinfo{author}{\bibfnamefont{M.~W.} \bibnamefont{Walser}},
  \bibinfo{author}{\bibfnamefont{C.}~\bibnamefont{Szymanowski}},
  \bibnamefont{and} \bibinfo{author}{\bibfnamefont{C.~H.}
  \bibnamefont{Keitel}}, \bibinfo{journal}{EPL (Europhysics Letters)}
  \textbf{\bibinfo{volume}{48}}, \bibinfo{pages}{533} (\bibinfo{year}{1999}).

\bibitem[{\citenamefont{Walser and Keitel}(2000)}]{Walser_2000b}
\bibinfo{author}{\bibfnamefont{M.~W.} \bibnamefont{Walser}} \bibnamefont{and}
  \bibinfo{author}{\bibfnamefont{C.~H.} \bibnamefont{Keitel}},
  \bibinfo{journal}{J. Phys. B} \textbf{\bibinfo{volume}{33}},
  \bibinfo{pages}{L221} (\bibinfo{year}{2000}).

\bibitem[{\citenamefont{Panek et~al.}(2002)\citenamefont{Panek,
  Kami\ifmmode~\acute{n}\else \'{n}\fi{}ski, and Ehlotzky}}]{Panek_2002}
\bibinfo{author}{\bibfnamefont{P.}~\bibnamefont{Panek}},
  \bibinfo{author}{\bibfnamefont{J.~Z.}
  \bibnamefont{Kami\ifmmode~\acute{n}\else \'{n}\fi{}ski}}, \bibnamefont{and}
  \bibinfo{author}{\bibfnamefont{F.}~\bibnamefont{Ehlotzky}},
  \bibinfo{journal}{Phys. Rev. A} \textbf{\bibinfo{volume}{65}},
  \bibinfo{pages}{033408} (\bibinfo{year}{2002}).

\bibitem[{\citenamefont{Panek et~al.}(2004)\citenamefont{Panek,
  Kami\ifmmode~\acute{n}\else \'{n}\fi{}ski, and Ehlotzky}}]{Panek_2004}
\bibinfo{author}{\bibfnamefont{P.}~\bibnamefont{Panek}},
  \bibinfo{author}{\bibfnamefont{J.~Z.}
  \bibnamefont{Kami\ifmmode~\acute{n}\else \'{n}\fi{}ski}}, \bibnamefont{and}
  \bibinfo{author}{\bibfnamefont{F.}~\bibnamefont{Ehlotzky}},
  \bibinfo{journal}{Phys. Rev. A} \textbf{\bibinfo{volume}{69}},
  \bibinfo{pages}{013404} (\bibinfo{year}{2004}).

\bibitem[{\citenamefont{Ahrens et~al.}(2012)\citenamefont{Ahrens, Bauke,
  Keitel, and M\"uller}}]{Ahrens_2012}
\bibinfo{author}{\bibfnamefont{S.}~\bibnamefont{Ahrens}},
  \bibinfo{author}{\bibfnamefont{H.}~\bibnamefont{Bauke}},
  \bibinfo{author}{\bibfnamefont{C.~H.} \bibnamefont{Keitel}},
  \bibnamefont{and} \bibinfo{author}{\bibfnamefont{C.}~\bibnamefont{M\"uller}},
  \bibinfo{journal}{Phys. Rev. Lett.} \textbf{\bibinfo{volume}{109}},
  \bibinfo{pages}{043601} (\bibinfo{year}{2012}).

\bibitem[{\citenamefont{Di~Piazza et~al.}(2010)\citenamefont{Di~Piazza,
  Milstein, and M\"uller}}]{DiPiazza_2010}
\bibinfo{author}{\bibfnamefont{A.}~\bibnamefont{Di~Piazza}},
  \bibinfo{author}{\bibfnamefont{A.~I.} \bibnamefont{Milstein}},
  \bibnamefont{and} \bibinfo{author}{\bibfnamefont{C.}~\bibnamefont{M\"uller}},
  \bibinfo{journal}{Phys. Rev. A} \textbf{\bibinfo{volume}{82}},
  \bibinfo{pages}{062110} (\bibinfo{year}{2010}).

\bibitem[{\citenamefont{M\"uller and M\"uller}(2011)}]{Muller_2011}
\bibinfo{author}{\bibfnamefont{T.-O.} \bibnamefont{M\"uller}} \bibnamefont{and}
  \bibinfo{author}{\bibfnamefont{C.}~\bibnamefont{M\"uller}},
  \bibinfo{journal}{Phys. Lett. B} \textbf{\bibinfo{volume}{696}},
  \bibinfo{pages}{201 } (\bibinfo{year}{2011}).

\bibitem[{\citenamefont{M\"uller and M\"uller}(2012)}]{Muller_2012}
\bibinfo{author}{\bibfnamefont{T.-O.} \bibnamefont{M\"uller}} \bibnamefont{and}
  \bibinfo{author}{\bibfnamefont{C.}~\bibnamefont{M\"uller}},
  \bibinfo{journal}{Phys. Rev. A} \textbf{\bibinfo{volume}{86}},
  \bibinfo{pages}{022109} (\bibinfo{year}{2012}).

\bibitem[{\citenamefont{Hu and Keitel}(1999)}]{Hu_1999}
\bibinfo{author}{\bibfnamefont{S.~X.} \bibnamefont{Hu}} \bibnamefont{and}
  \bibinfo{author}{\bibfnamefont{C.~H.} \bibnamefont{Keitel}},
  \bibinfo{journal}{Phys. Rev. Lett.} \textbf{\bibinfo{volume}{83}},
  \bibinfo{pages}{4709} (\bibinfo{year}{1999}).

\bibitem[{\citenamefont{Hu and Keitel}(2001)}]{Hu_2001}
\bibinfo{author}{\bibfnamefont{S.~X.} \bibnamefont{Hu}} \bibnamefont{and}
  \bibinfo{author}{\bibfnamefont{C.~H.} \bibnamefont{Keitel}},
  \bibinfo{journal}{Phy. Rev. A} \textbf{\bibinfo{volume}{63}},
  \bibinfo{pages}{053402} (\bibinfo{year}{2001}).

\bibitem[{\citenamefont{Walser and Keitel}(2001)}]{Walser_2001}
\bibinfo{author}{\bibfnamefont{M.~W.} \bibnamefont{Walser}} \bibnamefont{and}
  \bibinfo{author}{\bibfnamefont{C.~H.} \bibnamefont{Keitel}},
  \bibinfo{journal}{Opt. Commun.} \textbf{\bibinfo{volume}{199}},
  \bibinfo{pages}{447 } (\bibinfo{year}{2001}).

\bibitem[{\citenamefont{Walser et~al.}(2002)\citenamefont{Walser, Urbach,
  Hatsagortsyan, Hu, and Keitel}}]{Walser_2002}
\bibinfo{author}{\bibfnamefont{M.~W.} \bibnamefont{Walser}},
  \bibinfo{author}{\bibfnamefont{D.~J.} \bibnamefont{Urbach}},
  \bibinfo{author}{\bibfnamefont{K.~Z.} \bibnamefont{Hatsagortsyan}},
  \bibinfo{author}{\bibfnamefont{S.~X.} \bibnamefont{Hu}}, \bibnamefont{and}
  \bibinfo{author}{\bibfnamefont{C.~H.} \bibnamefont{Keitel}},
  \bibinfo{journal}{Phys. Rev. A} \textbf{\bibinfo{volume}{65}},
  \bibinfo{pages}{043410} (\bibinfo{year}{2002}).

\bibitem[{\citenamefont{Bhattacharyya et~al.}(2007)\citenamefont{Bhattacharyya,
  M., Chakrabarti, and Faisal}}]{Bhattacharyya_2007}
\bibinfo{author}{\bibfnamefont{S.}~\bibnamefont{Bhattacharyya}},
  \bibinfo{author}{\bibfnamefont{M.}~\bibnamefont{M.}},
  \bibinfo{author}{\bibfnamefont{J.}~\bibnamefont{Chakrabarti}},
  \bibnamefont{and} \bibinfo{author}{\bibfnamefont{F.~H.~M.}
  \bibnamefont{Faisal}}, \bibinfo{journal}{J. Phys. Conf. Ser.}
  \textbf{\bibinfo{volume}{80}}, \bibinfo{pages}{012029}
  (\bibinfo{year}{2007}).

\bibitem[{\citenamefont{Bhattacharyya et~al.}(2011)\citenamefont{Bhattacharyya,
  Mazumder, Chakrabarti, and Faisal}}]{Bhattacharyya_2011}
\bibinfo{author}{\bibfnamefont{S.}~\bibnamefont{Bhattacharyya}},
  \bibinfo{author}{\bibfnamefont{M.}~\bibnamefont{Mazumder}},
  \bibinfo{author}{\bibfnamefont{J.}~\bibnamefont{Chakrabarti}},
  \bibnamefont{and} \bibinfo{author}{\bibfnamefont{F.~H.~M.}
  \bibnamefont{Faisal}}, \bibinfo{journal}{Phys. Rev. A}
  \textbf{\bibinfo{volume}{83}}, \bibinfo{pages}{043407}
  (\bibinfo{year}{2011}).

\bibitem[{\citenamefont{Faisal}(2007{\natexlab{a}})}]{Faisal_2007a}
\bibinfo{author}{\bibfnamefont{F.~H.~M.} \bibnamefont{Faisal}},
  \bibinfo{journal}{J. Phys. B} \textbf{\bibinfo{volume}{40}},
  \bibinfo{pages}{F145} (\bibinfo{year}{2007}{\natexlab{a}}).

\bibitem[{\citenamefont{Faisal}(2007{\natexlab{b}})}]{Faisal_2007b}
\bibinfo{author}{\bibfnamefont{F.~H.~M.} \bibnamefont{Faisal}},
  \bibinfo{journal}{Phys. Rev. A} \textbf{\bibinfo{volume}{75}},
  \bibinfo{pages}{063412} (\bibinfo{year}{2007}{\natexlab{b}}).

\bibitem[{\citenamefont{Bargmann et~al.}(1959)\citenamefont{Bargmann, Michel,
  and Telegdi}}]{Bargmann_1959}
\bibinfo{author}{\bibfnamefont{V.}~\bibnamefont{Bargmann}},
  \bibinfo{author}{\bibfnamefont{L.}~\bibnamefont{Michel}}, \bibnamefont{and}
  \bibinfo{author}{\bibfnamefont{V.~L.} \bibnamefont{Telegdi}},
  \bibinfo{journal}{Phys. Rev. Lett.} \textbf{\bibinfo{volume}{2}},
  \bibinfo{pages}{435} (\bibinfo{year}{1959}).

\bibitem[{\citenamefont{Gersten and Mittleman}(1975)}]{Gersten_1975}
\bibinfo{author}{\bibfnamefont{J.~I.} \bibnamefont{Gersten}} \bibnamefont{and}
  \bibinfo{author}{\bibfnamefont{M.~H.} \bibnamefont{Mittleman}},
  \bibinfo{journal}{Phys. Rev. A} \textbf{\bibinfo{volume}{12}},
  \bibinfo{pages}{1840} (\bibinfo{year}{1975}).

\bibitem[{\citenamefont{Avetissian et~al.}(1997)\citenamefont{Avetissian,
  Markossian, Mkrtchian, and Movsissian}}]{Avetissian_1997}
\bibinfo{author}{\bibfnamefont{H.~K.} \bibnamefont{Avetissian}},
  \bibinfo{author}{\bibfnamefont{A.~G.} \bibnamefont{Markossian}},
  \bibinfo{author}{\bibfnamefont{G.~F.} \bibnamefont{Mkrtchian}},
  \bibnamefont{and} \bibinfo{author}{\bibfnamefont{S.~V.}
  \bibnamefont{Movsissian}}, \bibinfo{journal}{Phys. Rev. A}
  \textbf{\bibinfo{volume}{56}}, \bibinfo{pages}{4905} (\bibinfo{year}{1997}).

\bibitem[{\citenamefont{Krainov}(1997)}]{Krainov_1997}
\bibinfo{author}{\bibfnamefont{V.~P.} \bibnamefont{Krainov}},
  \bibinfo{journal}{J. Opt. Soc. Am. B} \textbf{\bibinfo{volume}{14}},
  \bibinfo{pages}{425} (\bibinfo{year}{1997}).

\bibitem[{\citenamefont{Avetissian et~al.}(1999)\citenamefont{Avetissian,
  Hatsagortsian, Markossian, and Movsissian}}]{Avetissian_1999}
\bibinfo{author}{\bibfnamefont{H.~K.} \bibnamefont{Avetissian}},
  \bibinfo{author}{\bibfnamefont{K.~Z.} \bibnamefont{Hatsagortsian}},
  \bibinfo{author}{\bibfnamefont{A.~G.} \bibnamefont{Markossian}},
  \bibnamefont{and} \bibinfo{author}{\bibfnamefont{S.~V.}
  \bibnamefont{Movsissian}}, \bibinfo{journal}{Phys. Rev. A}
  \textbf{\bibinfo{volume}{59}}, \bibinfo{pages}{549} (\bibinfo{year}{1999}).

\bibitem[{\citenamefont{Smirnova et~al.}(2008)\citenamefont{Smirnova, Spanner,
  and Ivanov}}]{Smirnova_2008}
\bibinfo{author}{\bibfnamefont{O.}~\bibnamefont{Smirnova}},
  \bibinfo{author}{\bibfnamefont{M.}~\bibnamefont{Spanner}}, \bibnamefont{and}
  \bibinfo{author}{\bibfnamefont{M.}~\bibnamefont{Ivanov}},
  \bibinfo{journal}{Phys. Rev. A} \textbf{\bibinfo{volume}{77}},
  \bibinfo{pages}{033407} (\bibinfo{year}{2008}).

\bibitem[{\citenamefont{Klaiber
  et~al.}(2013{\natexlab{a}})\citenamefont{Klaiber, Yakaboylu, and
  Hatsagortsyan}}]{Klaiber_2013a}
\bibinfo{author}{\bibfnamefont{M.}~\bibnamefont{Klaiber}},
  \bibinfo{author}{\bibfnamefont{E.}~\bibnamefont{Yakaboylu}},
  \bibnamefont{and} \bibinfo{author}{\bibfnamefont{K.~Z.}
  \bibnamefont{Hatsagortsyan}}, \bibinfo{journal}{Phys. Rev. A}
  \textbf{\bibinfo{volume}{87}}, \bibinfo{pages}{023417}
  (\bibinfo{year}{2013}{\natexlab{a}}).

\bibitem[{\citenamefont{Klaiber
  et~al.}(2013{\natexlab{b}})\citenamefont{Klaiber, Yakaboylu, and
  Hatsagortsyan}}]{Klaiber_2013b}
\bibinfo{author}{\bibfnamefont{M.}~\bibnamefont{Klaiber}},
  \bibinfo{author}{\bibfnamefont{E.}~\bibnamefont{Yakaboylu}},
  \bibnamefont{and} \bibinfo{author}{\bibfnamefont{K.~Z.}
  \bibnamefont{Hatsagortsyan}}, \bibinfo{journal}{Phys. Rev. A}
  \textbf{\bibinfo{volume}{87}}, \bibinfo{pages}{023418}
  (\bibinfo{year}{2013}{\natexlab{b}}).

\bibitem[{\citenamefont{Popov}(2004)}]{Popov_2004u}
\bibinfo{author}{\bibfnamefont{V.~S.} \bibnamefont{Popov}},
  \bibinfo{journal}{Phys. Usp.} \textbf{\bibinfo{volume}{47}},
  \bibinfo{pages}{855} (\bibinfo{year}{2004}).

\bibitem[{\citenamefont{Klaiber
  et~al.}(2013{\natexlab{c}})\citenamefont{Klaiber, Yakaboylu, Bauke,
  Hatsagortsyan, and Keitel}}]{Klaiber_2013c}
\bibinfo{author}{\bibfnamefont{M.}~\bibnamefont{Klaiber}},
  \bibinfo{author}{\bibfnamefont{E.}~\bibnamefont{Yakaboylu}},
  \bibinfo{author}{\bibfnamefont{H.}~\bibnamefont{Bauke}},
  \bibinfo{author}{\bibfnamefont{K.~Z.} \bibnamefont{Hatsagortsyan}},
  \bibnamefont{and} \bibinfo{author}{\bibfnamefont{C.~H.}
  \bibnamefont{Keitel}}, \bibinfo{journal}{Phys. Rev. Lett.}
  \textbf{\bibinfo{volume}{110}}, \bibinfo{pages}{153004}
  (\bibinfo{year}{2013}{\natexlab{c}}).

\end{thebibliography}

\end{document}